\documentclass{cas-dc}

\usepackage{amsmath}
\usepackage{hyperref}
\usepackage[numbers]{natbib}
\begin{document}
\emergencystretch 3em

%\begin{frontmatter}

\title[mode=title]{Determination of angular distributions from the high efficiency solenoidal separator SOLITAIRE}

\shorttitle{}

\author[]{L.~T.~Bezzina}%[orcid=0000-0002-5718-8230]
\cormark[1] 
%\ead{lauren.bezzina@anu.edu.au}

\address[]{Department of Nuclear Physics, Research School of Physics, Australian National University, Canberra 2601, Australia}

\author[]{E.~C.~Simpson}

\author[]{D.~J.~Hinde}

\author[]{M.~Dasgupta}

\author[]{I.~P.~Carter}

\author[]{D.~C.~Rafferty}

\cortext[cor1]{Corresponding author. Email: lauren.bezzina@anu.edu.au} 

\shortauthors{L.~T.~Bezzina et~al.}

%%%%%%%%%%%%%%%%%%%%%%%%%%%%%%%%%%%%%%%%%%%%%%%%%%
\begin{abstract}
A novel fusion product separator, based on a gas-filled 8~T superconducting solenoid has been developed at the Australian National University. Though the transmission efficiency of the solenoid is very high, precision cross section measurements require knowledge of the angular distribution of the evaporation residues. 

\noindent A method has been developed to deduce the angular distribution of the evaporation residues from the laboratory-frame velocity distribution of the evaporation residues measured at the exit of the separator. The features of this method are presented, focusing on the example of $^{34}$S+$^{89}$Y which is compared to an independent measurement of the angular distribution. The establishment of this method now allows the novel solenoidal separator to be used to obtain reliable, precision fusion cross-sections.

\end{abstract}


%\end{frontmatter}
\maketitle

%%%%%%%%%%%%%%%%%%%%%%%%%%%%%%%%%%%%%%%%%%%%%%%%%%
\section{Introduction}
\label{sec:intro}

Precise measurements of fusion cross sections are key to advancing our understanding of heavy-ion fusion \cite{dasgupta_arnp_1998,back_RMP_2013}.  They have revealed the role of nuclear structure in enhancing sub-barrier fusion cross sections \cite{dasso_npa_1983b,stokstad_prc_1980} and, via measurements of barrier distributions \cite{rowley_PLB_1991}, demonstrated the importance of deformations \cite{leigh_prc_1993} and surface vibrations \cite{stefanini_prl_1995}.  As the field seeks to understand the observed above-barrier suppression of fusion \cite{newton_prc_2004}, hindrance in deep sub-barrier fusion \cite{back_RMP_2013}, and efforts continue to synthesise new super-heavy elements to map out the so-called ``island of stability'', precise fusion cross sections will continue to be critical.

Experimental determination of fusion cross sections involves measurement of the product nuclei generated following the formation of a fused intermediate, the compound nucleus (CN). Two main decays modes are observed: fission into two fragments, or emission of nucleons (or clusters of nucleons) to form an evaporation residue (ER). The balance between these decay modes depends on the properties of the compound nucleus formed, namely, mass, charge, excitation energy and angular momentum.

The measurement of ER cross sections is particularly challenging. Momentum conservation leads to an ER angular distribution that is strongly forward-focused, but the cross section is, at best, $10^4$ times smaller than elastic scattering at the same angles.  The major challenge is then the efficient and effective separation of the evaporation residues from the elastically scattered beam particles. For barrier distribution measurements, which require cross sections with uncertainties of 1\% or better, it is vital that the transmission efficiency of the separator be known very well. At the Department of Nuclear Physics at the Australian National University (ANU), separation is achieved using a separation system based around a gas-filled superconducting solenoid, called SOLITAIRE \cite{rodriguez_nima_2010}. In this paper, we describe a method for deducing the angular distribution of the ERs from the measured velocity distribution, and use this information to deduce the transport efficiency of the device and thus determine precise ER cross sections. 

This method was initially presented at the conference Fusion17 and published in the proceedings of that conference~\cite{bezzina_fusion_2017}. While the method remains similar in form, this paper presents it with greater detail and tests it with greater rigour, instilling greater confidence in the use of this method to determine precise ER cross sections.

%%%%%%%%%%%%%%%%%%%%%%%%%%%%%%%%%%%%%%%%%%%%%%%%%%
\section{SOLITAIRE}
\label{sec:solitaire}

The superconducting \underline{sol}enoid for \underline{i}n-beam \underline{t}ransport \underline{a}nd \underline{i}dentification of \underline{r}ecoiling \underline{e}vaporation-residues (SOLITAIRE) separates evaporation residues from the intense flux of elastically scattered beam particles using the magnetic field of the solenoid. The principles of operation and first measurements of SOLITAIRE are detailed by Rodr\'{i}guez~\cite{rodriguez_nima_2010}, and as such, only a brief summary follows. 

\begin{figure*}
    \centering
    \includegraphics[width=0.8\textwidth]{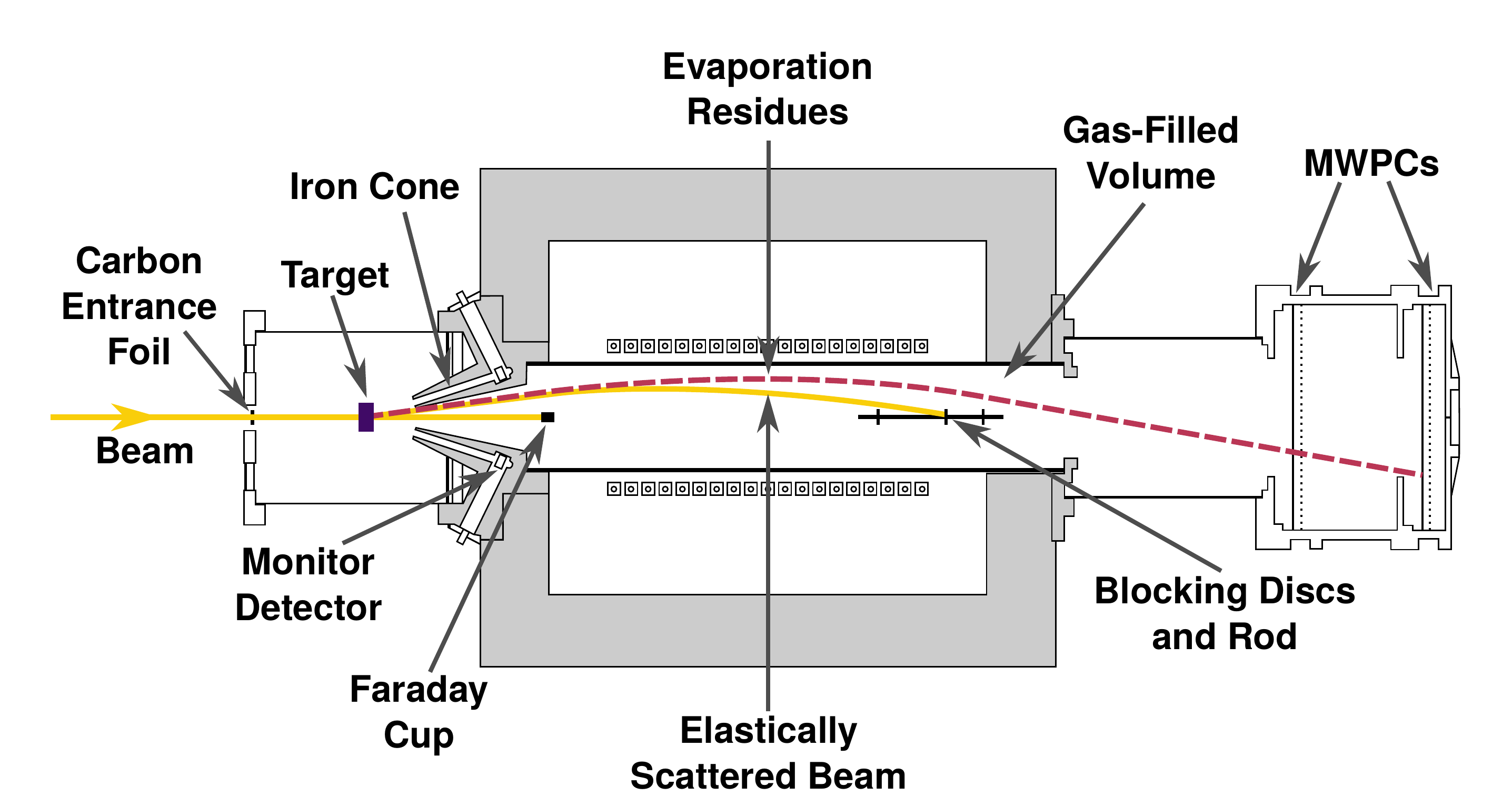}
    \caption{Schematic diagram of SOLITAIRE. This figure is adapted from \cite{rodriguez_nima_2010}, and shows key components of the device, as well as indicative radial trajectories of evaporation residues (purple dashed line) and elastically scattered beam particles (orange line). The shaded grey area indicates the iron shielding of the solenoid, including a `nose' cone at the solenoid entrance. Also indicated are two of the four monitor detectors used for normalisation to elastic scattering, and the two multiwire proportional counters (MWPCs) used for direct detection of evaporation residues.}
    \label{fig:solitaire}
\end{figure*}

The 14UD tandem accelerator at the ANU Heavy Ion Accelerator Facility provides pulsed beams of the desired energy, which are transported to the SOLITAIRE target chamber (as shown in Figure \ref{fig:solitaire}).  The target chamber contains four silicon monitor detectors placed at $18^{\circ}$, which are used for normalisation purposes.  Immediately prior to the target chamber the beam passes through a thin carbon foil, which separates the helium gas-filled region from the upstream beam line.  The energy losses of the primary beam in both the carbon foil and in the helium gas prior to the reaction target can be significant, particularly at near-barrier energies, and must be taken into account.  This is achieved by scattering the beam from a thin gold foil in the target position and measuring the change in energy in the monitor detectors with and without the helium gas and carbon foil present. For the benchmarking reactions discussed in this paper the energy loss quantified in this way was $0.92\pm0.01$~MeV.

Following the reaction target, an on-axis Faraday Cup blocks the unscattered beam and any particles with laboratory scattering angles $\theta_l<0.5^{\circ}$.  The iron nose cone of the solenoid blocks all particles with $\theta_l>9^{\circ}$. ERs and elastically scattered beam ions within these angular limits proceed into the solenoid.

The solenoid acts as a thin converging lens for the charged particles, separating them based on their magnetic rigidity \cite{rodriguez_nima_2010, klemperer_1953}. Upon entering the solenoid the ERs and elastically scattered beam have the same average momentum and often have overlapping charge state distributions, and therefore similar rigidity. For this reason, the solenoid bore is filled with a low pressure (usually $\sim$ 1~torr) helium gas, subjecting the ions passing through it to charge-changing collisions. The resulting average charge state, $\left\langle q \right\rangle$, is dependent on the velocity and atomic number of the nuclei \cite{betz_rmp_1972}. The significantly higher velocity of the elastically scattered beam nuclei bring them to a higher charge state than the lower velocity ERs.

With a higher charge state, the elastically scattered particles have a lower rigidity and shorter focal length, and are intercepted by the blocking rod and discs placed along the solenoid axis in the solenoid bore. The lower charge-state ERs exit the solenoid, and are brought to a focus further downstream and are detected in SOLITAIRE's two multi-wire proportional counters (MWPCs). The MWPCs record position, energy loss, and timing information of each event. 

Owing to the large solid angle of acceptance of the solenoid (86~msr \cite{rodriguez_nima_2010}) the transmission efficiency of the device is very high. This means SOLITAIRE has the capability to measure ER cross sections with very high precision. In order to accurately extract ER cross sections, however, a number of quantities need to be precisely known:

\begin{enumerate}
\item the angular distribution of ERs leaving the target,
    \item the fraction of ERs entering the solenoid,
    \item the fraction of ERs which are transported through to the detectors,
    \item the fraction of ERs incident on the detectors which are recorded in the data acquisition system, and
    \item beam normalisation through Rutherford scattering at the monitor detectors.
\end{enumerate}{}

Throughout this paper, points 2. and 3. are combined to define a `transport efficiency', that is, the ratio of ERs exiting the solenoid relative to those produced at the target. The primary factor affecting the transport efficiency is the angular acceptance, defined by the Faraday Cup and iron nose cone.  The probability that any given event with laboratory angle $\theta_l$ is transported through the solenoid can be simulated using Monte Carlo techniques, but the total transport efficiency requires knowledge of the angular distribution of ERs (point 1). This angular distribution must be deduced from the experimental measurements, as, while statistical model calculations using packages such as PACE4 \cite{gavron_prc_1980} are possible in principle, the input parameters of these calculations are not constrained enough to precisely predict experimental angular distributions. 

The complication in the present work is that any measurement of the ERs takes place after the solenoid, and is therefore already moderated by the transport efficiency. This paper specifically addresses this conundrum, and presents a method to deduce both the transport efficiency of the device and the angular distribution of ERs, as described in the next section.

%%%%%%%%%%%%%%%%%%%%%%%%%%%%%%%%%%%%%%%%%%%%%%%%%%
\section{Method}
\label{sec:method}

We now discuss the method for extracting the transport efficiency of the solenoid and ER angular distribution from the target. As the MWPCs are position sensitive, one might hope to use the position information to directly reconstruct the initial angular distribution of the evaporation residues.  However, this approach is sensitive to the details of the scattering and charge-changing interactions with the helium fill gas, which spread the trajectories of the transmitted ions. This also requires precise knowledge of the focal point of the ERs \cite{carter_hias_2012}.  A new and more reliable approach has been concieved, to use the velocity distribution of the ERs. The velocity itself can be measured using the timing of the pulsed beam and the MWPC signals. The velocity is unaffected on average by the interactions with the gas, and is therefore representative of the velocity distribution of the ERs at the reaction target, albeit filtered by transport through the solenoid.

\begin{figure}
    \centering
    \includegraphics[width=0.48\textwidth]{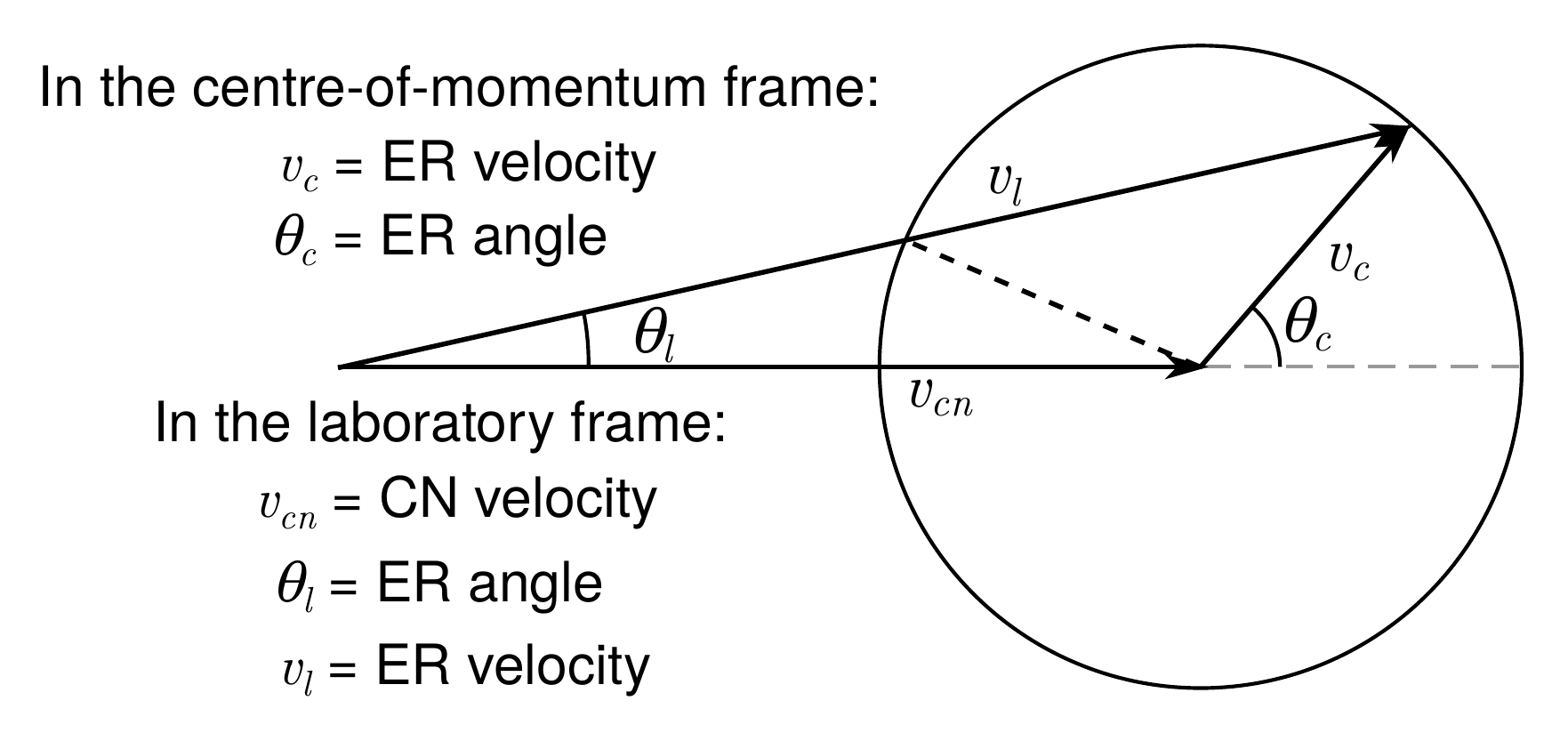}
    \caption{Velocity vector diagram showing the relationship between velocities of compound nucleus (CN) and evaporation residue (ER) in centre-of-momentum and laboratory frames. The compound nucleus velocity in the laboratory frame is $v_{cn}$, which is related to the ER velocity and angle in the laboratory frame ($v_l$ and $\theta_l$, respectively) via the ER velocity and angle in the centre-of-momentum frame of the compound nucleus ($v_c$ and $\theta_c$).}
    \label{fig:velocityvectors}
\end{figure}

The laboratory velocity and angular distributions of ERs are linked via the distribution of events in the centre-of-momentum frame, demonstrated by the velocity vector diagram shown in Figure \ref{fig:velocityvectors}. This diagram illustrates the net recoil of an evaporation residue following the emission of nucleons or nucleon clusters from the compound nucleus. The final ER then has some velocity and angle relative to the beam axis in the centre-of-momentum frame, labelled $v_c$ and $\theta_c$, respectively. Summed with the velocity of the compound nucleus in the laboratory frame, $v_{cn}$, the ER has resulting laboratory velocity and angle ($v_l$, $\theta_l$), which are experimentally measurable quantities.

If one assumes some form for the probability of ER recoils in the centre-of-momentum frame, that is, the distribution of events with ($v_c,\theta_c$) any cross section (e.g., $d\sigma/d\theta_l$, $d\sigma/dv_l$, $d\sigma/d\Omega$) can be reconstructed.  Here, we write this in terms of the unit normalised recoil velocity probability distribution $P(v_c)$, and its angular distribution $W(\theta_c)$. To demonstrate this relationship between $P(v_c)$ and $\sigma_{ER}(v_l)$ we present the double differential cross section \cite{dick_ejp_2009}:

\begin{samepage}
    \begin{align}
        \frac{d^2\sigma_{ER}}{dv_l\;dv_c} = 2\pi \sigma_{ER} \left[\frac{v_l}{v_c\;v_{cn}}\right]  W(\theta_c) P(v_c) \;,
        \label{eqn:vlvc}
    \end{align}
\end{samepage}

\noindent where $\sigma_{ER}$ is the total evaporation residue cross section and all other variables are as defined above and in Figure \ref{fig:velocityvectors}. For the rest of this paper, we assume the net emission (including all $xn$, $pxn$ and $\alpha xn$ channels) is isotropic, as tests of the sensitivity of our results to an expected range of anisotropies in $W(\theta_c)$~(using an upper limit determined from~\cite{bencyjohn_prc_1997}) had negligible effects on the shape and normalisation of the final distributions. The procedure used to extract $\sigma_{ER}$ and $P(v_c)$ from the velocity measurements made at the MWPCs is described below.

%%%%%%%%%%%%%%%%%%%%%%%%%%%%%%%%%%%%%%%%%%%%%%%%%%
\subsection{Evaporation residue transport through the solenoid}
\label{subsec:MC}

The first component required is an estimation of the transport efficiency of the solenoid for each ER with a given laboratory velocity ($v_l$) and angle ($\theta_l$).  Note that there is a distinction between the transport efficiency for a particular $\theta_l$ and $v_l$, denoted $\varepsilon(\theta_l,v_l)$, and the \emph{total} transport efficiency $\varepsilon_T$, which is integrated over all velocities and angles which are emitted from the target. The calculations for $\varepsilon(\theta_l,v_l)$ were performed with the Monte Carlo program Solix \cite{solix}, developed at the ANU. The program simulates the passage of particles of specified mass, initial energy and angle through the solenoid. As well as accounting for the physical obstructions of the blocking discs and rod (among others), it models the charge-changing collisions with the specified fill gas, and can count the number of particles incident on any specified area along the solenoid's axis.

The ERs were simulated as originating from a beam with a Gaussian profile with a width of $\sigma=0.5$~mm in both $x$ and $y$ (perpendicular to the $z$ of the beam axis) though it was found that the resulting efficiency map was largely insensitive to the precise size of the simulated beam spot. The range of velocities was chosen to correspond to the measured velocity distribution range, while the range of angles slightly exceeded the angular acceptance of the iron nose cone of the solenoid. The mean charge state of the ERs was assumed to depend linearly on the velocity \cite{betz_rmp_1972}. While more complex models of the charge state’s velocity dependence exist~\cite{betz_rmp_1972,gregorich_prc_2005}, the average charge state in all of these models is very nearly linear in velocity, as noted in Refs.~\cite{gregorich_prc_2005,rodriguez_phd_2009}. 

\begin{figure}
    \centering
    \includegraphics[width=0.48\textwidth]{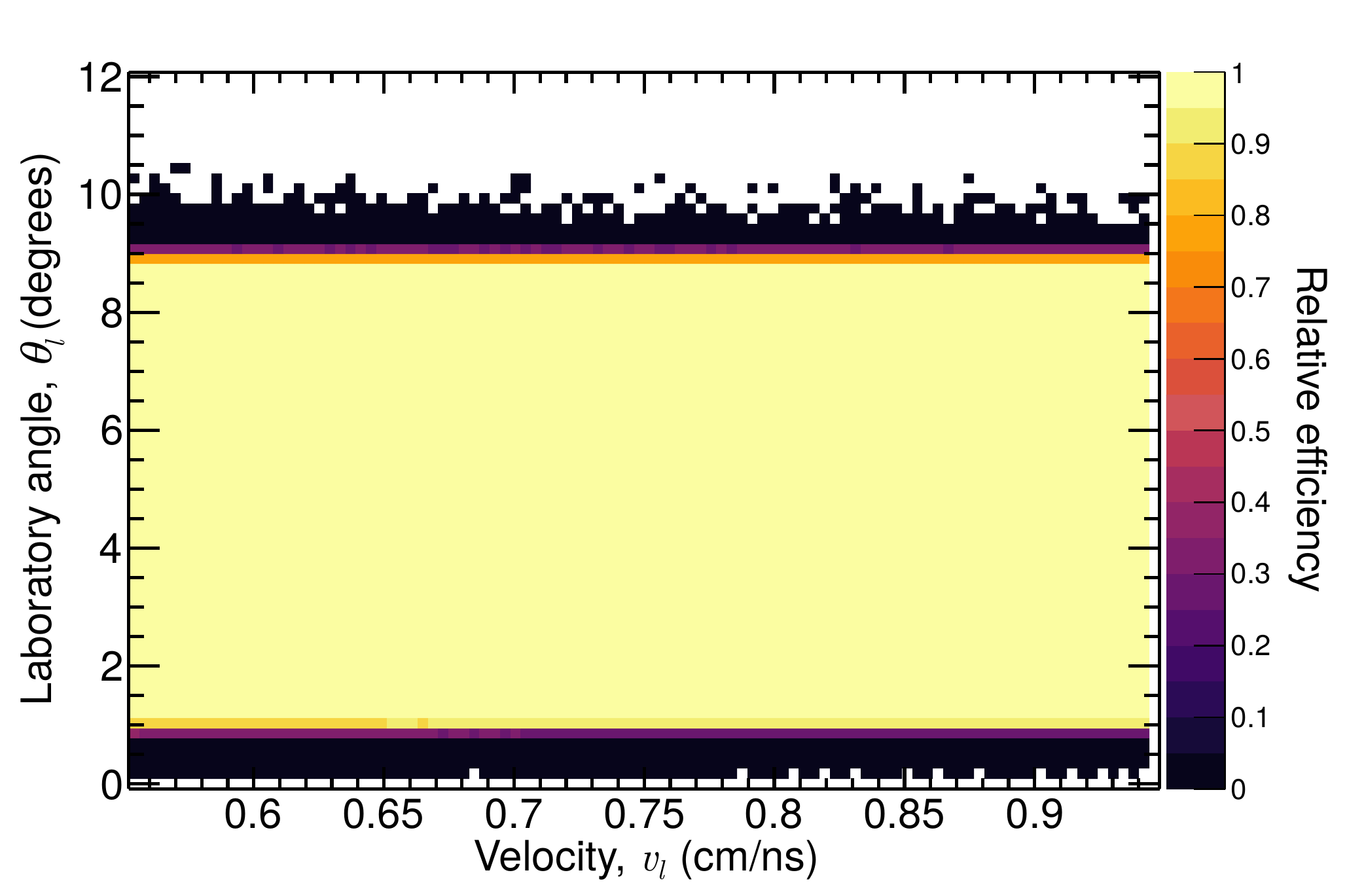}
    \includegraphics[width=0.48\textwidth]{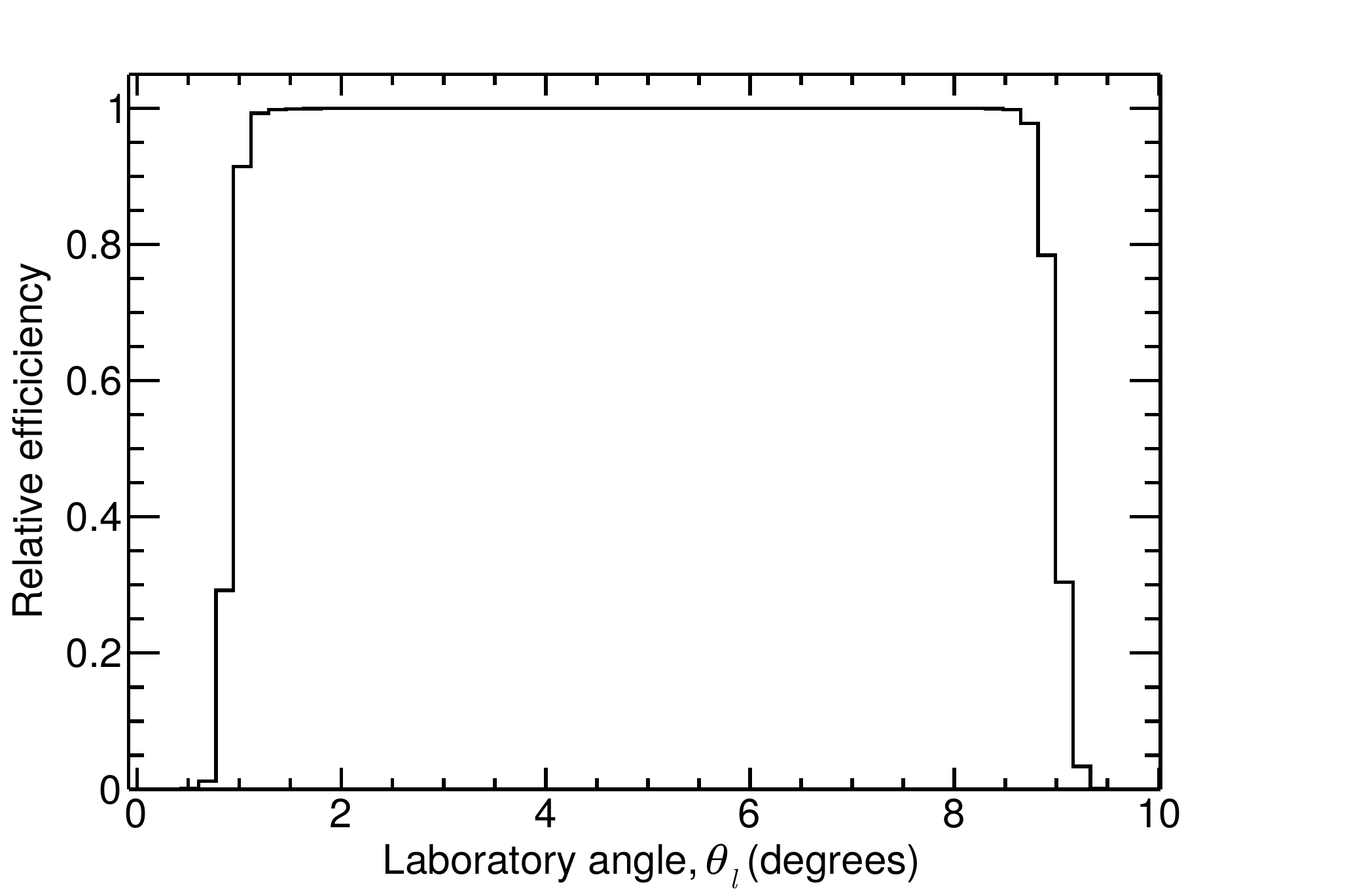}
    \caption{The two-dimensional efficiency map (top), $\varepsilon(\theta_l,v_l)$, of SOLITAIRE for the $^{34}$S+$^{89}$Y reaction, is uniform in velocity $v_l$, but has a strong dependence on the laboratory angle, $\theta_l$. The bottom figure shows the efficiency as a function of laboratory angle only, in order to more clearly demonstrate the angular dependence of the efficiency.}
    \label{fig:efficiency}
\end{figure}

The simulated transport efficiency for the ERs from the $^{34}$S+$^{89}$Y reaction is presented in Figure~\ref{fig:efficiency}, with the central velocity corresponding to that appropriate for $E_{lab}=124$ MeV.  Due to the assumption that the mean charge state is proportional to the velocity, the ion rigidity and thus efficiency are independent of velocity.  In angle, the Faraday cup stops all ions (direct beam and ERs) with $\theta_l<0.5^{\circ}$.  The cut at $\theta_l>9^{\circ}$ is due to the angular acceptance of the iron cone of the solenoid.

%%%%%%%%%%%%%%%%%%%%%%%%%%%%%%%%%%%%%%%%%%%%%%%%%%
\subsection{The form of $P(v_c)$}
\label{subsec:pvc}

With the transport efficiency $\varepsilon(\theta_l,v_l)$ simulated, we then need to deduce the form of the ER recoil velocity distribution in the compound nucleus frame, $P(v_c)$. As outlined by Weisskopf \cite{weisskopf_pr_1937}, the distribution of ER velocities in the compound-nucleus rest-frame is expected to take the shape of one or more Maxwellian distributions. This expectation arises from a statistical description of the compound nucleus (CN) due to the high level-density of the accessible excited states. Different modes of evaporation (e.g. $n$, $p$, $\alpha$) have a differing energy threshold for emission, which is dependent on the Coulomb barrier between the emitted particle and remaining residue. For each evaporation residue, a series of emissions is likely to have occurred. 

These factors lead us to expect $P(v_c)$ to be composed of several Maxwellian functions, one with no offset from $v_c=0$, representative of $xn$ emission, where there is no barrier to emission. There could also be Maxwellians representing $pxn$, $\alpha xn$, $\alpha pxn$ and so on, each of which would have been offset from zero proportional to the magnitude of the Coulomb barrier faced by the charged particle/s emitted. 

In our case, the compound-nucleus rest frame distribution is parameterized using the sum of two Maxwellians, the first representing neutron and proton emission, while the second simulated $\alpha xn$-emission (and therefore being offset from zero). The initial parameters were estimated using a a bootstrap extraction of $P(v_c)$ from $\sigma_{ER(exp)(v_l)}$, following a method outlined in \cite{carter_hias_2012}, based on Equation~\ref{eqn:vlvc}. More complex forms for $P(v_c)$ are discussed in Section \ref{subsec:angdist}.

%%%%%%%%%%%%%%%%%%%%%%%%%%%%%%%%%%%%%%%%%%%%%%%%%%
\subsection{Iterative Correction Procedure}
\label{subsec:iterative}

Having chosen a parameterised form of $P(v_c)$, we then use a correction procedure to deduce the total transport efficiency of the solenoid $\varepsilon_T$ and therefore the evaporation residue angular distribution.  First, the timing information (corrected for energy loss in the He fill gas) from the front MWPC is used to produce a distribution of events in laboratory velocity $v_l$. This experimental distribution, measured at the MWPCs, is filtered by the SOLITAIRE transport efficiency. Next, taking the assumed form for $P(v_c)$, a theoretical double differential cross section $d^2\sigma_{ER}/d\theta_ldv_l$ is generated using: 

\begin{equation}
    \frac{d^2\sigma_{ER}}{d\theta_l\;dv_l} = 2\pi \sigma_{ER} \sin{\theta_l} W(\theta_c) P(v_c) \frac{v_l^2}{v_c^2}\;.
\label{eqn:tlvl}
\end{equation}

\begin{figure}
    \centering
    \includegraphics[width=0.50\textwidth]{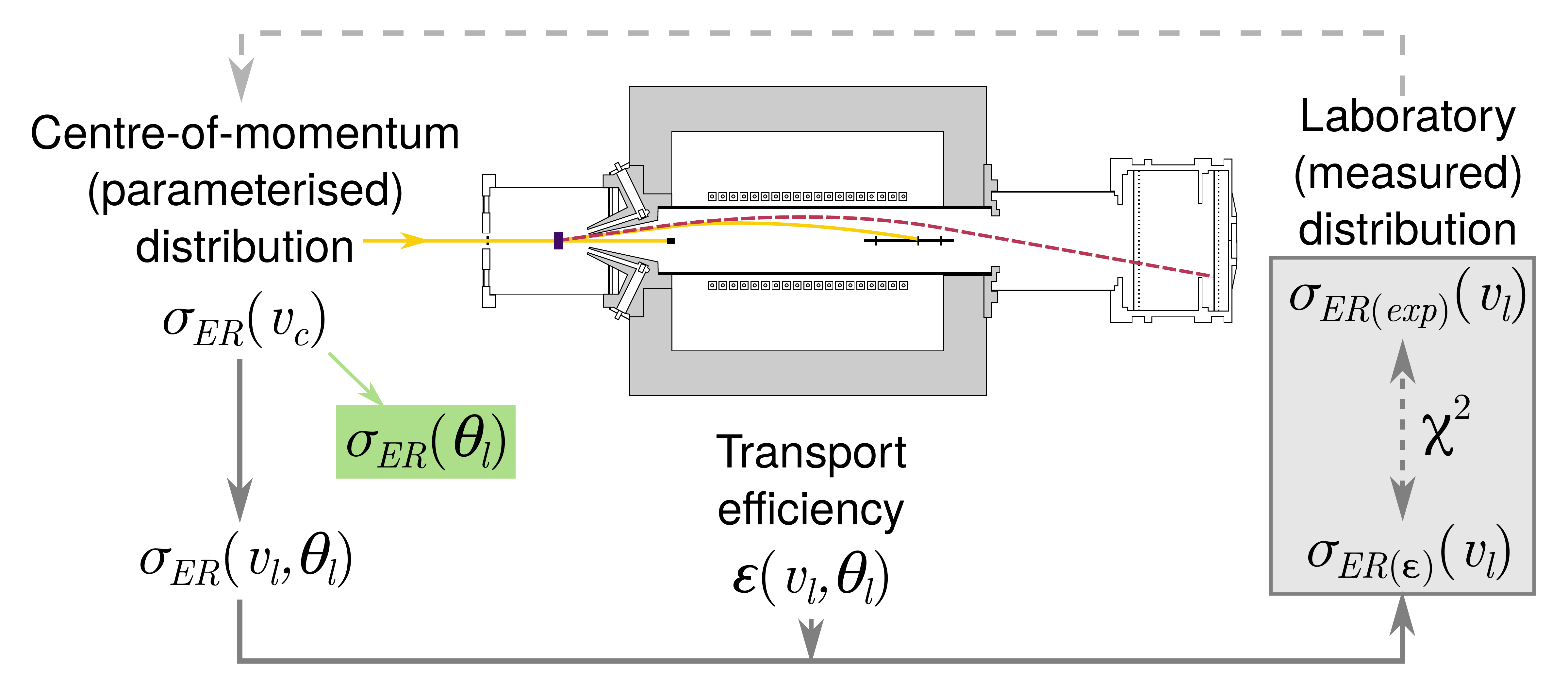}
    \caption {Flowchart of the method for extracting precise angular distributions from measurements made with the SOLITAIRE device, with a minimised schematic of the device included for context. The flowchart shows the routine described in Section~\ref{sec:method}, including the initialisation step for estimating initial parameters of the $\sigma_{ER}(v_c)$ distribution, and the final transformation to the angular distribution $\sigma_{ER}(\theta_l)$.}
    \label{fig:flowchart}
\end{figure}

\begin{figure}
    \centering
    \includegraphics[width=0.48\textwidth]{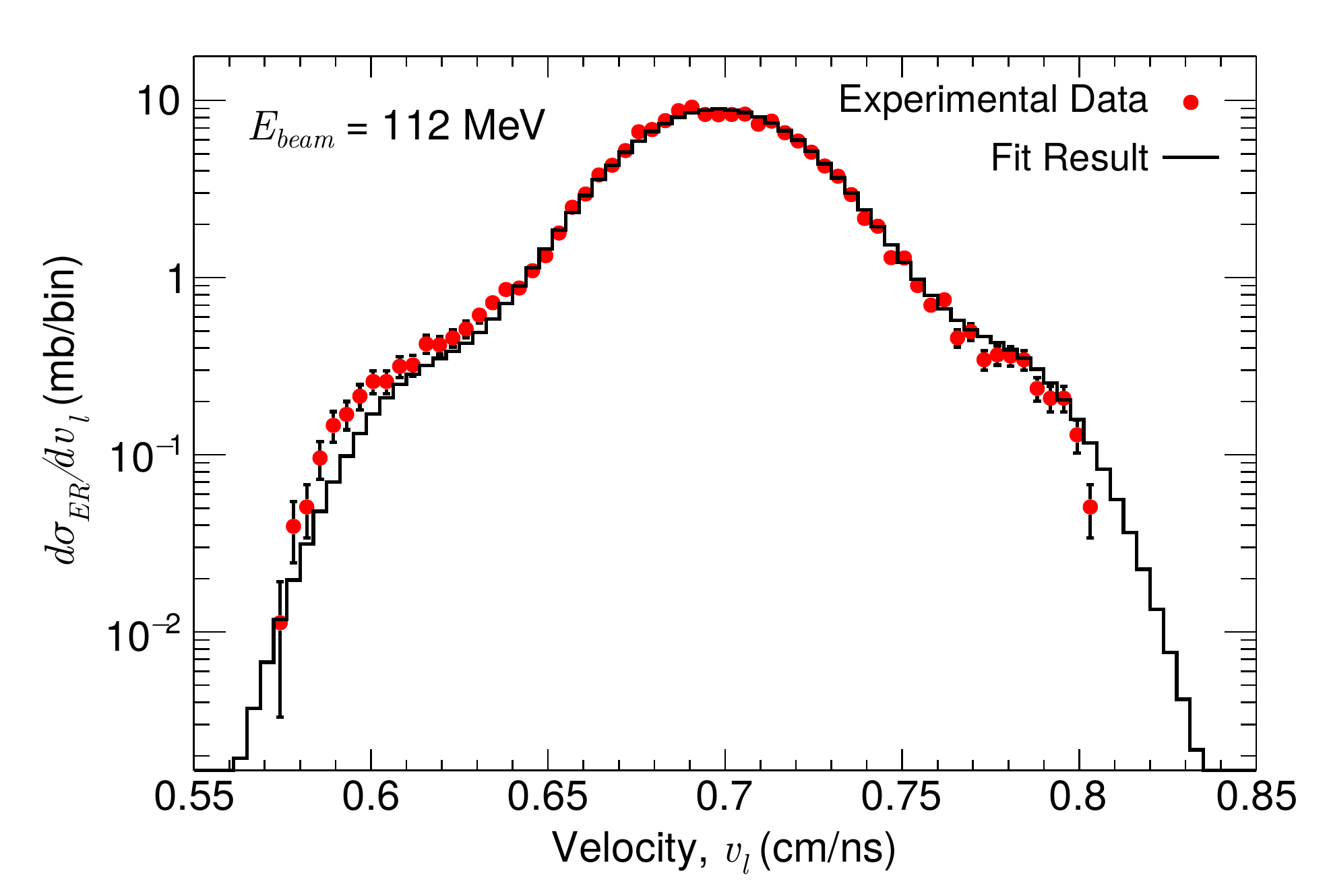}
    \includegraphics[width=0.48\textwidth]{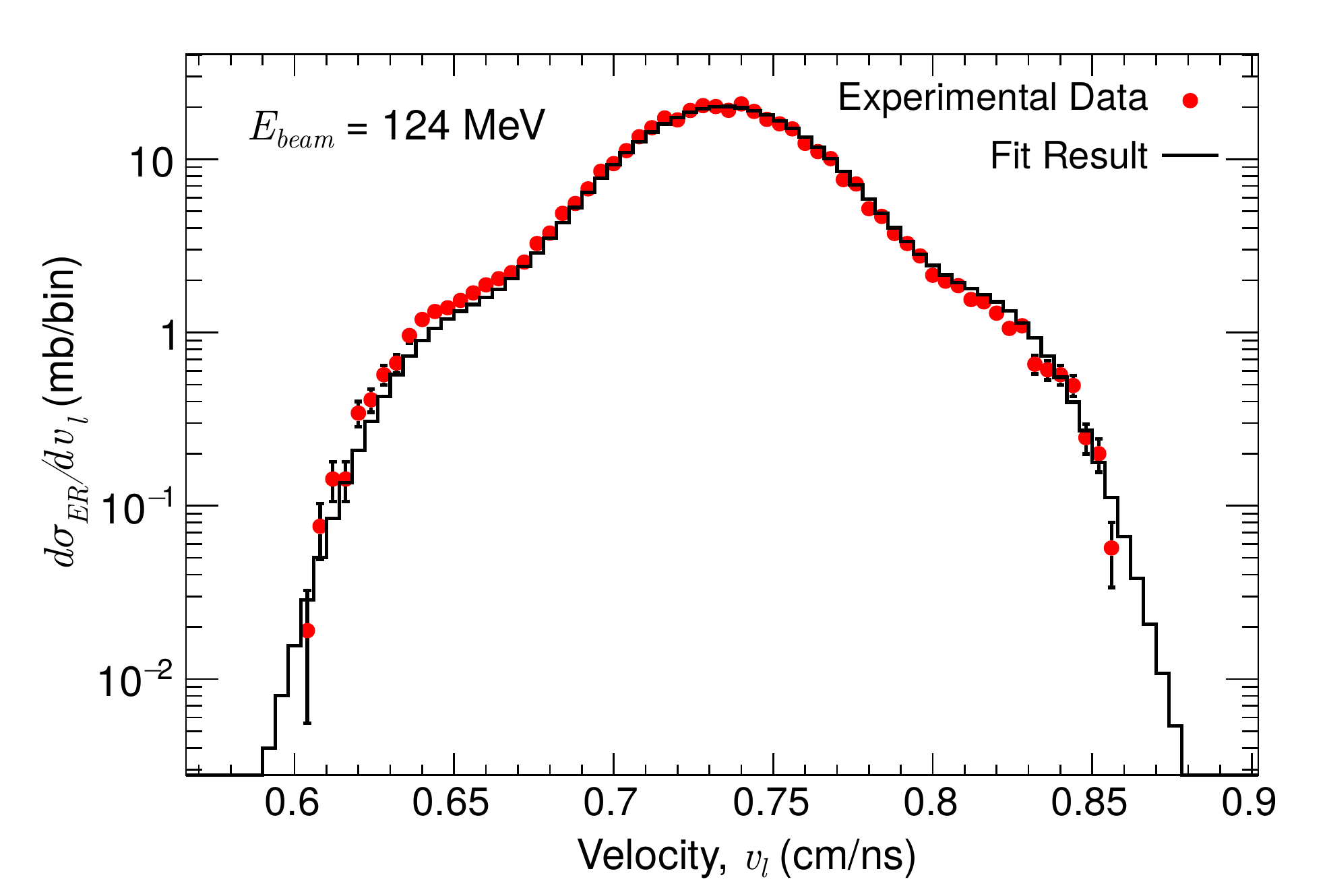}
    \caption {A comparison of the laboratory-frame velocity distributions at the beam energies $E_{beam}=112$~MeV (top) and $E_{beam}=124$~MeV (bottom). The experimental data, $\sigma_{ER(exp)}(v_l)$, are shown as red points, while the result of the minimisation routine, $\sigma_{ER(\varepsilon)}(v_l)$, is shown as a black histogram. The experimental distribution is naturally filtered by the SOLITAIRE transport efficiency, while the result of the minimisation routine is filtered by the simulated transport efficiency.}
    \label{fig:velocitycomparison}
\end{figure}

This represents the distribution of ERs emitted from the reaction target, which then needs to be filtered by the transport efficiency in order to compare to the measured laboratory velocity distribution. The cross section $d^2\sigma_{ER}/d\theta_ldv_l$ is simply multiplied by the transport efficiency $\varepsilon(\theta_l,v_l)$ bin-by-bin.  Integrating the result over the laboratory angle $\theta_l$ then gives an efficiency-filtered velocity distribution, $\sigma_{ER(\varepsilon)}(v_l)$, which allows a direct comparison with the laboratory velocity distribution measured at the exit of the solenoid, $\sigma_{ER(exp)}(v_l)$, and evaluation of the $\chi^2$ of these two distributions. This series of transformations, beginning from the creation of the double differential cross section $d^2\sigma_{ER}/d\theta_ldv_l$, is then optimised by iterating over the parameters of $P(v_c)$ in order to minimise the $\chi^2$. The procedure is illustrated in Figure~\ref{fig:flowchart}. Examples of the final, optimised velocity distribution for $E_{beam}=112$ and 124 MeV are shown alongside the experimentally measured velocity distributions in Figure \ref{fig:velocitycomparison}. In both cases the agreement is very good.

    %%%%%%%%%%%%%%%%%%%%%%%%%%%%%%%%%%%%%%%%%%%%%%%%%%
\section{Results}
\label{sec:results}

\subsection{Optimised $P(v_c)$}
\label{subsec:optpvc}

\begin{figure}
    \centering
    \includegraphics[width=0.48\textwidth]{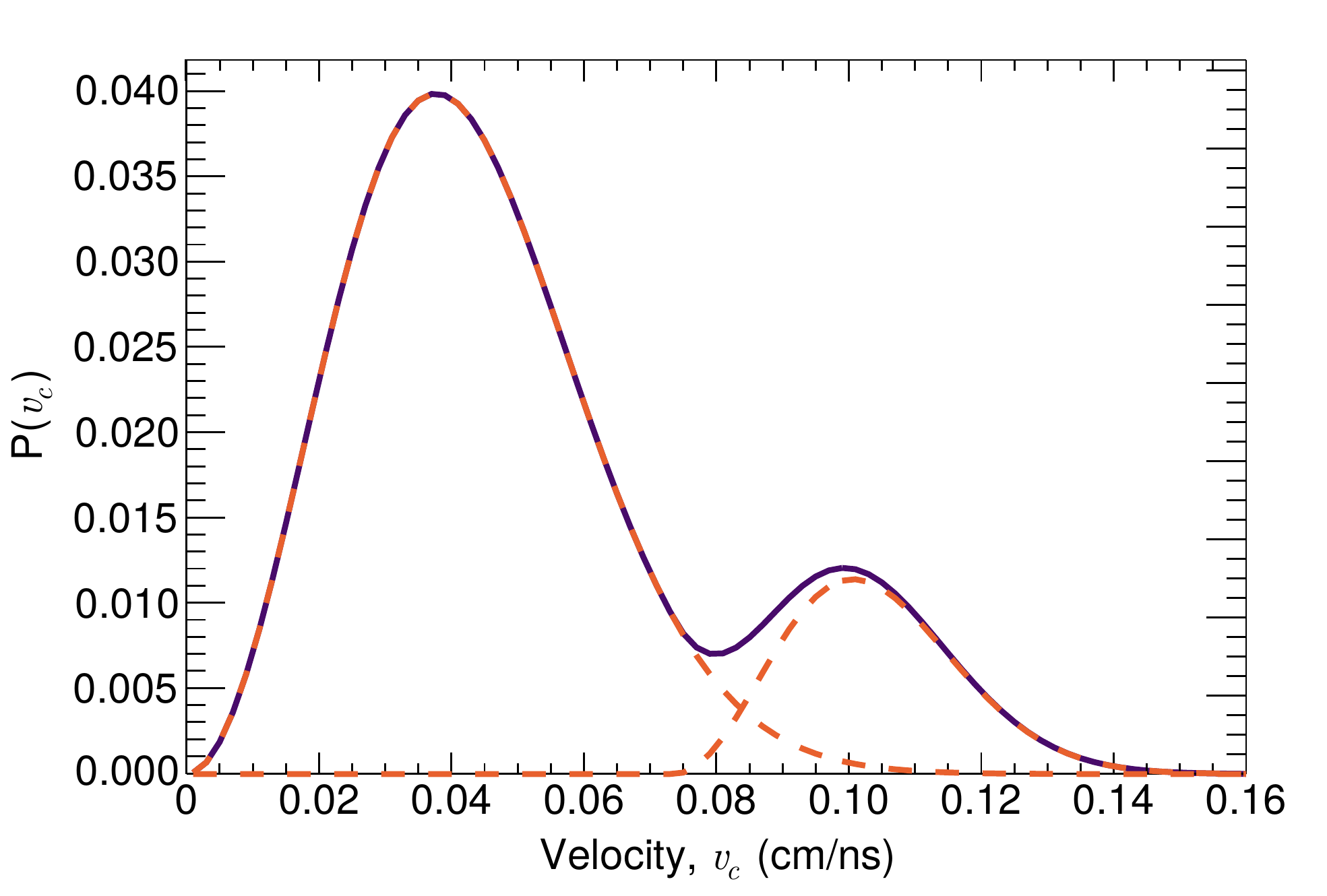}
    \caption {The optimised centre-of-momentum frame velocity distribution for the case of $E_{beam}=124$~MeV. The solid line shows the total ER distribution, including all emission modes. The dashed lines show the two component ER distributions, one peaked at $v_c\sim0.037$~cm/ns representing ERs from $xn$ and $pxn$ emission, while the other represents ERs resulting from $\alpha xn$ emission.}
    \label{fig:comvelocity}
\end{figure}

The first confirmation of the suitability of our procedure comes from an examination of the final, optimised $P(v_c)$. While the offset of the first Maxwellian was fixed at 0~cm/ns (expected to represent $xn$ and $pxn$ emission), no other parameters were fixed or limited. The expected value of the offset of the second Maxwellian physically corresponds to the minimum recoil velocity an ER can be expected to have, assuming a single alpha is emitted. In the scenario with the lowest centre-of-momentum energy, this corresponds to the barrier energy between the alpha particle and resultant ER, which is then shared between the two nuclei in proportion to their masses. Using a S{\~a}o Paulo potential for the nuclear part of the barrier potential, the expected offset for $\alpha$ emission from the CN was then calculated to be 0.0642~cm/ns. The iterative correction routine finds an optimised offset of $0.073\pm0.002$~cm/ns, which is in reasonable agreement with the calculated value. The final, optimised $P(v_c)$ distribution is shown in Figure~\ref{fig:comvelocity}.

%%%%%%%%%%%%%%%%%%%%%%%%%%%%%%%%%%%%%%%%%%%%%%%%%%
\subsection{Angular distributions}
\label{subsec:angdist}

\begin{figure}
    \centering
    \includegraphics[width=0.48\textwidth]{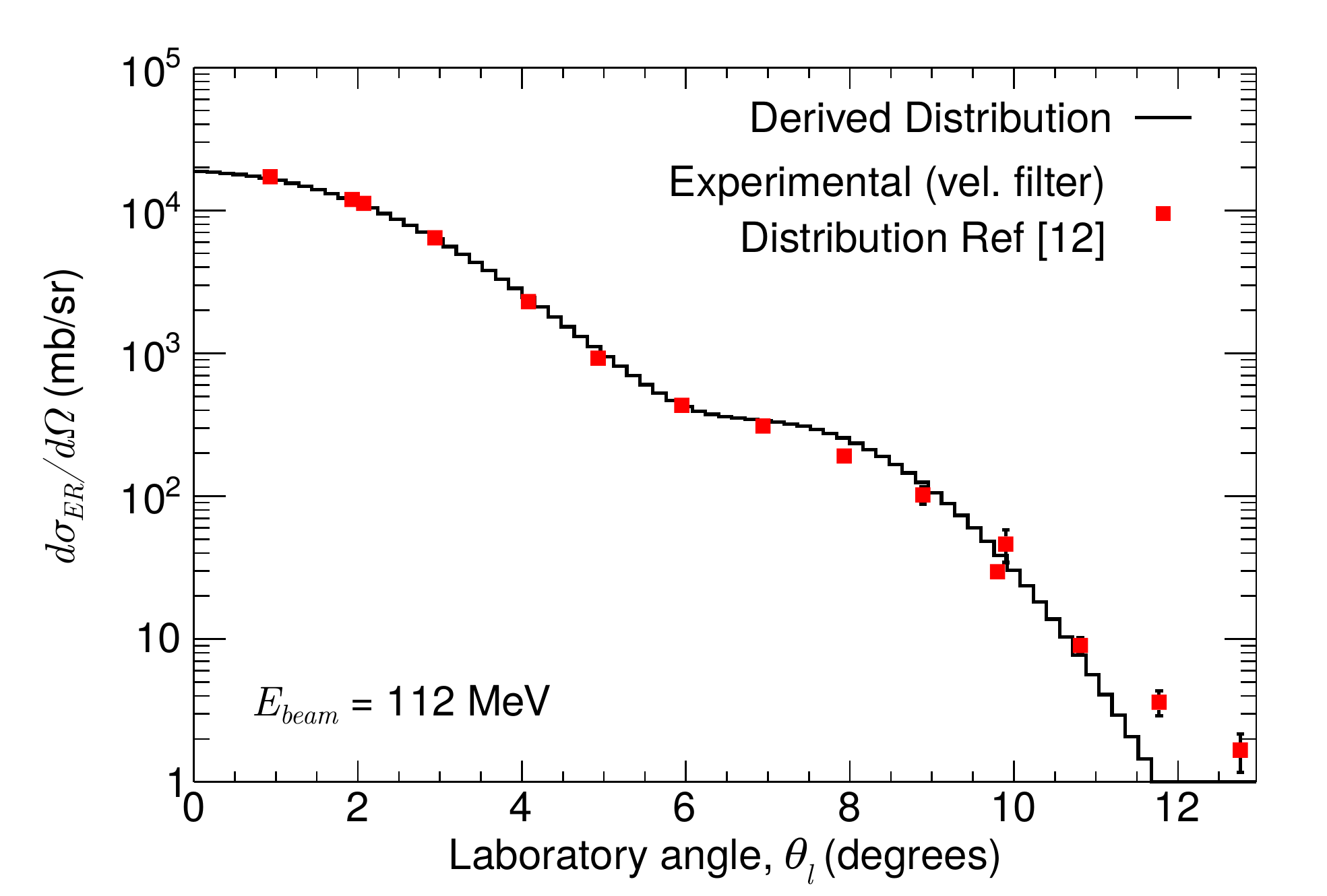}
    \includegraphics[width=0.48\textwidth]{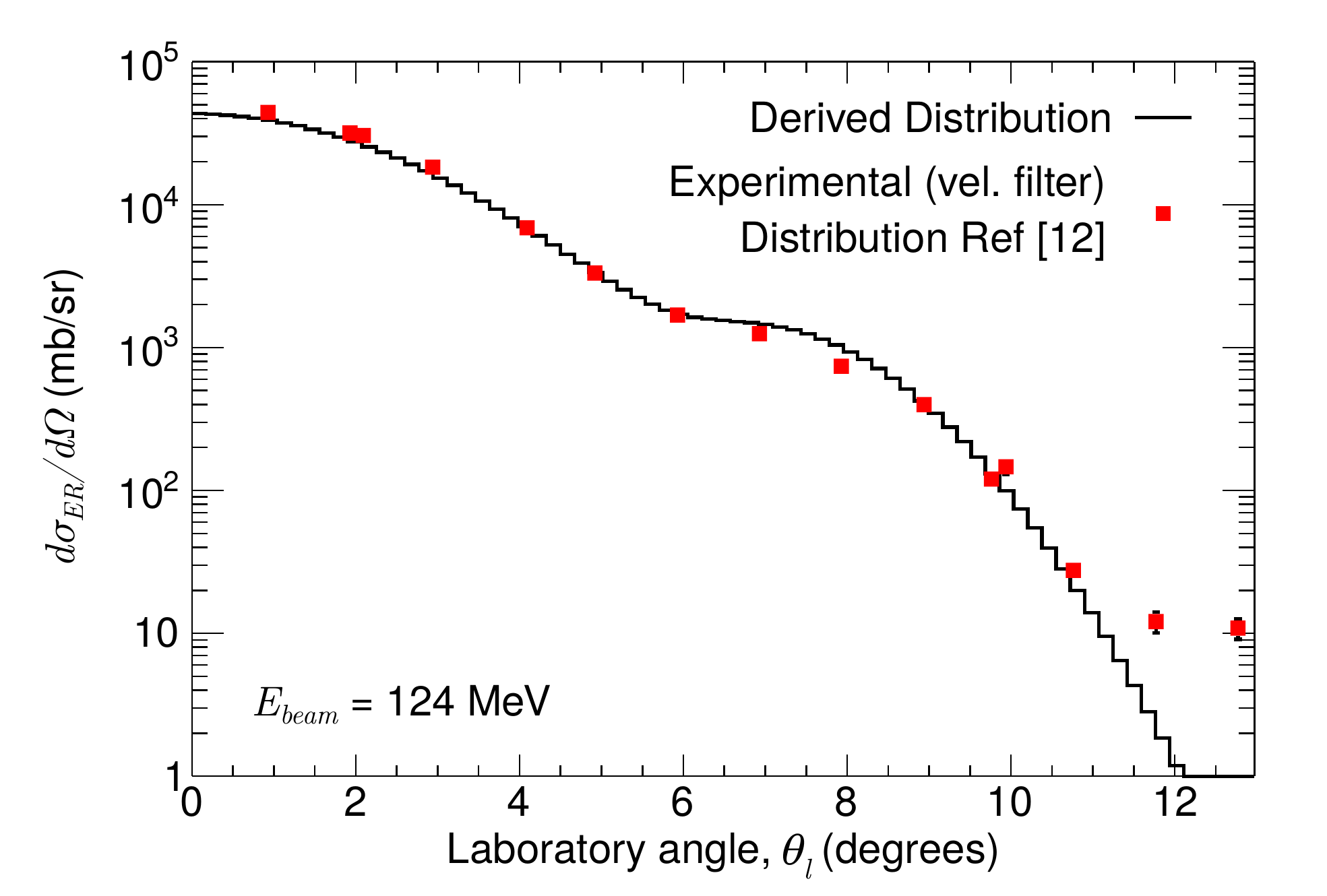}
    \caption{The differential cross section $d\sigma_{ER}/d\Omega$ as a function of laboratory angle, $\theta_l$, for $E_{beam}=112$ MeV (top) and $E_{beam}=124$ MeV (bottom). The result of the minimisation routine is shown as a black histogram, while independent experimental data \cite{mukherjee_prc_2002} is shown in red squares.}
    \label{fig:comparison}
\end{figure}

We next compare results for the extracted angular distribution to the independent velocity filter measurements of Ref.~\cite{mukherjee_prc_2002}. Due to energy losses in the carbon foil and helium gas as discussed in Section~\ref{sec:solitaire} above, the energies of the collisions differ slightly ($\sim0.3$ MeV) from those of Ref.~\cite{mukherjee_prc_2002}. As it is not expected that this small difference will significantly alter the shape of the angular distributions, we have interpolated the cross sections of Ref.~\cite{mukherjee_prc_2002} to the mid-target energies and scaled the experimental angular distributions of Ref.~\cite{mukherjee_prc_2002} for comparison. The two distributions are presented together in Figure \ref{fig:comparison} for both energies.  

Excellent agreement is seen between both distributions. The two-shouldered angular distribution arises from distinct contributions where only proton and neutron evaporation occurs, and when $\alpha$ evaporation occurs.  This structure is very well reproduced, particularly in the relative magnitudes of the two components, and the position of the secondary shoulder of the distribution between $8-10^{\circ}$.

One of the considerable advantages of using SOLITAIRE is apparent when we consider that the velocity filter measurements (made angle-by-angle) took a number of days to acquire in full, limited mostly by the time taken to collect enough counts at the largest angles of the distribution. The measurements with SOLITAIRE, on the other hand, took only tens of minutes.  The overall transport efficiency of SOLITAIRE can be found by comparing the integral of $d\sigma_{ER}/d\theta_ldv_l$ before and after filtering by the SOLITAIRE transport efficiency. For the $^{34}$S+$^{89}$Y reaction at both beam energies, the total transport efficiency, $\varepsilon_T$, is 92\%. 

The largest discrepancy occurs for the $E_{beam}=124$~MeV measurements at forward angles, where the cross section is underestimated. This deviation may be a significant reason for the discrepancy in the total cross section, discussed in the next section. At these forward angles, due to blocking by the Faraday cup, the transport efficiency is particularly low.  In terms of the centre-of-momentum velocity distribution, the largest contributions to these small angles comes from ERs with small recoil velocities $v_c$, and it may be that this portion of $P(v_c)$ is (relatively) poorly constrained by our iterative procedure.

Whilst the angular distributions are reasonably well described using two Maxwellians in $P(v_c)$, it may be that a sum of three Maxwellians (describing $xn$, $pxn$ and $\alpha xn$ emission explicitly) would be more appropriate, with the $pxn$ component, like the $\alpha xn$ component, having an offset.  Currently, the lower Maxwellian is forced to describe both $xn$ and $pxn$ emissions, which might lead to an underestimate of small $v_c$.  We might expect this problem to be more pronounced at $E_{beam}=124$~MeV, where proton emission is more likely, which is consistent with the poorer description of the angular distribution at this energy.  The fact that, at both energies, the angular distribution shows a smoother transition between the two shoulders for the velocity filter data, also indicates that the ER recoil velocity distribution may be more complex than assumed.  The inclusion of a third Maxwellian would certainly smooth the extracted distribution in this region.  Attempts were made to fit the velocity distribution with three Maxwellians, but there are insufficient counts in the present data to unambiguously constrain the parameters, and further measurements are required to clarify this point.

A deviation from the independent measurement is also present at larger angles ($\theta_l\geq12^{\circ}$), where the deduced distributions underestimate the independent experimental data at both energies. This may be due to the presence of another emission mode, $\alpha+pxn$/$^{7}$Li+$xn$, which would deliver greater recoil to the ERs than the $\alpha$ emission, resulting in more widely distributed ERs. Attempts to fit an additional Maxwellian failed, this time due to inadequate constraint in the extremes of the velocity distribution which would uniquely constrain the Maxwellian contributing to these largest angles.

Despite these discrepancies, the method as it currently stands is well-suited for measuring ER cross sections in systems where neutron evaporation is the dominant emission mode. In such systems, we can expect that the emission is described well by two Maxwellians in the centre-of-momentum frame.

%%%%%%%%%%%%%%%%%%%%%%%%%%%%%%%%%%%%%%%%%%%%%%%%%%

\subsection{Total cross sections}
\label{subsec:xsectotal}

With respect to cross sections, we first consider the total evaporation residue cross sections for the incident beam energies $E_{beam}=112$ and 124~MeV, comparing to those of Ref. \cite{mukherjee_prc_2002}, measured using a velocity filter technique.  Due to the slightly differing energies, we have interpolated the results of \cite{mukherjee_prc_2002} to match our mid-target energies.  Taking into account measured energy losses in the entrance window carbon foil, the helium gas prior to the target, and half the target thickness, our mid-target energy corresponding to $E_{beam}=112$~MeV is $E_{lab}=111.18$~MeV, and the extracted evaporation residue cross section is $\sigma_{ER} = 158\pm1$~mb, where the limiting error is the statistical error.  This is in excellent agreement with the value of $158\pm1$~mb from Ref. \cite{mukherjee_prc_2002}.  For the $E_{beam}=124$~MeV runs, the mid-target energy is $E_{lab}=123.17$~MeV, and the reconstructed cross section is $424\pm2$~mb.  This compares to an interpolated cross section of $453\pm5$~mb from Ref. \cite{mukherjee_prc_2002}.  Possible reasons for this discrepancy in the higher energy data are as discussed in the previous section. Further, as indicated in the list in Section~\ref{sec:solitaire}, a number of other quantities need to be precisely known in order to correctly normalise the experimental distribution, and this discrepancy requires further investigation.

%%%%%%%%%%%%%%%%%%%%%%%%%%%%%%%%%%%%%%%%%%%%%%%%%%
\subsection{Uncertainty in the method}
\label{subsec:bootstrap}

To quantify the uncertainty introduced by the reconstruction routine itself, a test was designed based on statistical bootstrapping \cite{efron_1979}.  Sub-samples of the measured data are taken and run through the routine to calculate the cross section $\sigma_{ER}$, exactly as would be done for the total data set. The data for these sub-samples is a random sample of events from the original data, allowing repeat sampling. 

By choosing the random sub-samples a number of times (100 has been shown to be sufficient \cite{goodhue_mis_2012}) a mean value of the cross section can be calculated, along with the standard error in the mean for each sub-sample size. The standard error in the mean is a purely statistical quantity, and estimates how far the mean of the sub-samples is likely to be from the population mean, renormalising the standard deviation of the sub-samples by $1/\sqrt n$, where $n$ is the number of sub-samples. It does not take any systematic error into account. This allows two important properties of the routine to be deduced: its consistency and its bias \cite{eadie_1971}. If the routine is consistent, it is expected that the mean value of the cross section should converge to the `true' value as the sub-sample size approaches the size of the data set. If the routine is \emph{unbiased}, then for any sub-sample size, the routine should always produce a value close to the `true' value, within statistical uncertainties.

\begin{figure}
    \centering
    \includegraphics[width=0.45\textwidth]{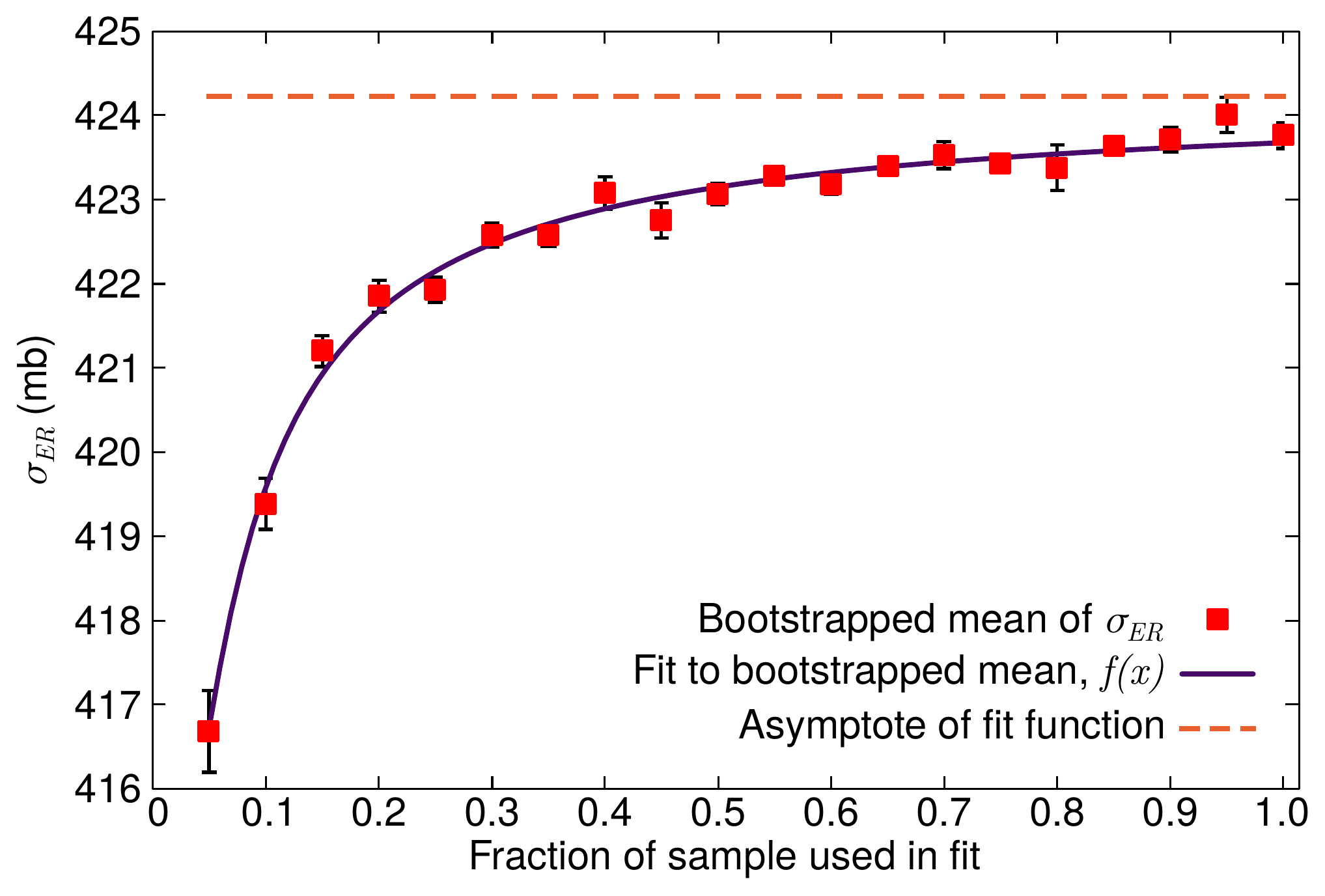}
    \caption{Results of the statistical bootstrap performed for the routine described in Section~\ref{sec:method}. The plot shows mean cross sections calculated from 100 sub-samples at each fractional size, with error bars representing the standard error in the mean, as well as a fit to these means and the asymptote of that fit.}
    \label{fig:bootstrap}
\end{figure}

The resulting mean cross section as a function of sample size is shown in Figure~\ref{fig:bootstrap}. The results show that the routine is biased at smaller sample sizes, but it is consistent, as it asymptotically converges on a value as the sample size increases (approaching a single value in the infinite limit~\cite{takeshi_1985}). The bias is likely introduced due to inadequate constraint of the second (offset) Maxwellian, since the sub samples tend to have insufficient data away from the central peak of the velocity distribution.  This suggestion is supported by examination of the behaviour of individual parameter values as a function of sub-sample size. There was some bias apparent in the parameter controlling the width of the first Maxwellian, becoming consistent at a sub-sample size of 30\%. The parameters controlling the width and offset of the second Maxwellian, however, showed variations an order of magnitude larger, and remained biased at sub-samples sizes up to 50\%. As the total cross section is also biased up to these sub-sample sizes, it is suggested that constraint of the second Maxwellian is only possible once 50\% of the data is used in the routine.

In order to test whether our reconstruction routine had satisfactorily converged to the correct answer, we fit the bootstrapped data with a set of converging functions. The chosen set of functions were of the form $c_0+c_1x^{-1}+...+c_nx^{-n}$ where the $c_i$ are fit parameters. The simplest function of this type with a reduced $\chi^2$ closest to 1 (the optimal fit) was $f(x) = c_0+c_1x^{-1}+c_2x^{-2}$, adding further terms offered no improvement to the reduced $\chi^2$, and so this relatively simple function was chosen. Once the fit was performed, the constant parameter $c_0$ could be extracted, which is the horizontal asymptote the function (and hence bootstrapped data) converges to in the infinite limit (as per our above definition of consistency). After fitting, it was found that $c_0 = 424.218 \pm 0.072$~mb.

We can then compare our bootstrapped means (red squares in Figure~\ref{fig:bootstrap}) to the asymptote. When we have enough data, we would expect that the bootstrapped mean would lie along the asymptotic line (dashed orange). Currently, the mean values using a sub-sample of 50\% or greater are within $\sim$~0.2\% of this asymptote (lower limit 423.37 mb), but as they do not agree with the line within error, it is possible that for this estimator, more data is needed to satisfactorily converge (i.e. within error) to a final value.

Characterisation of the routine in this way allows us to conclude that for the double Maxwellian case, we must have at least 50\% of the data points measured in the benchmarking experiment ($\sim$40000 ER counts) in order to use the routine without significant bias.  Assessment of the error bars representing the standard error in the mean for these results also allow us to propose that the uncertainty of the routine itself is of the order of 0.5~mb. Of the total 424~mb, this corresponds to 0.1\%, even less than the statistical uncertainty associated with a total sample of 40000 events, which corresponds to 0.5\%.  This method of analysing the routine itself will be invaluable as the suggested changes to the form of the $v_c$ distribution are made, allowing us to assess how well constrained a third Maxwellian may be to new datasets with larger numbers of counts. 

%%%%%%%%%%%%%%%%%%%%%%%%%%%%%%%%%%%%%%%%%%%%%%%%%%

\section{Summary and Outlook}
\label{sec:conclusions}

A new method to characterise the transport efficiency of the fusion product separator SOLITAIRE has been developed. The approach uses Monte Carlo simulations of the transport efficiency and measured ER velocity distributions to iteratively reconstruct the ER angular distribution and cross section.  The routine has been benchmarked against the reaction of $^{34}$S+$^{89}$Y at two beam energies: 112~MeV and 124~MeV. Good agreement is found between the deduced ER angular distributions and an independently measured angular distribution~\cite{mukherjee_prc_2002}.The cross sections for these reactions have been determined to be $158\pm1$~mb and $424\pm2$~mb, respectively, with a corresponding transport efficiency of 92\% for both cases.  The total cross section is in excellent agreement at $E_{beam}=112$ MeV. However, it is underestimated by $\sim7\%$ at $E_{beam}=124$ MeV, possibly due to the simplified form of the ER recoil distribution $P(v_c)$ used. This uncertainty will be explored using new cross section measurements made using an upgraded version of the SOLITAIRE devices, featuring a new 8T solenoid.

%%%%%%%%%%%%%%%%%%%%%%%%%%%%%%%%%%%%%%%%%%%%%%%%%%
\section*{Acknowledgements}
The project was supported by the Australian Research Council Discovery Grants FL110100098, DP160101254, DP170102318 and DP170102423. Support for operations of the HIAF accelerator facility from the Federal Government NCRIS HIA capability is also acknowledged.

%%%%%%%%%%%%%%%%%%%%%%%%%%%%%%%%%%%%%%%%%%%%%%%%%%
\bibliography{LBezzina_bibliography_NIMA}
%\bibliographystyle{cas-model2-names}

%%%%%%%%%%%%%%%%%%%%%%%%%%%%%%%%%%%%%%%%%%%%%%%%%%
\end{document}